\documentstyle[11pt,aaspp4]{article}
\tighten
\input psfig.tex

\slugcomment{Submitted to the Astrophysical Journal Letters}

\begin{document}

\title{Annihilation Fountain in the Galactic Center Region}

\author{C. D. Dermer \& J. G. Skibo}

\affil{E. O. Hulburt Center for Space Research, Code 
7653,\\ Naval Research Laboratory, Washington, DC 20375-5352}

\begin{abstract}

Two different model-independent mapping techniques have been applied to {\it
CGRO} OSSE, {\it SMM}, TGRS and balloon data and reveal a feature in the 0.511
MeV e$^+$-e$^-$ annihilation radiation pattern of our galaxy centered at $\ell
\sim -2^\circ $ and $b  \sim 10^\circ$ with a flux of $\sim 5 \cdot 10^{-4}$
0.511 MeV ph cm$^{-2}$ s$^{-1}$.  If near the galactic center, then e$^+$
sources are producing $\sim 10^{42}$ $e^+$ s$^{-1}$ which annihilate $\approx
1$-$2$ kpc above the galactic plane. A starburst episode within the inner few
hundred pc of our galaxy would drive hot pair-laden gas into the halo, with the
one-sidedness pointing to the site of initial pressure release at the onset of
the starburst activity.  Positrons lose energy and annihilate as they are
convected upward with the gas flow, and we calculate high-latitude annihilation
patterns and fluxes in accord with the observations. Changes in the ionization
state when the escaping gas cools could give annihilation radiation
substructure. The fountain of hot ($\sim 10^6$-$10^7$ K) gas rising into the
galactic halo would be seen through its enhanced dispersion measure, thermal
emission, and recombination radiation. 

\end{abstract}


\clearpage

\section{Introduction}

In addition to the previously identified (Purcell et al. 1993; Ramaty, Skibo \&
Lingenfelter 1994) 0.511 MeV annihilation glow from the disk and nuclear bulge
of our galaxy, recent analyses (Purcell et al. 1997a, 1997b; Cheng et al. 1997)
of data obtained with the Oriented Scintillation Spectrometer Experiment (OSSE)
on the {\it Compton Gamma Ray Observatory}, the {\it Solar Maximum Mission},
the Transient Gamma Ray Spectrometer (Teegarden et al. 1996) on the {\it Wind}
spacecraft, and GRIS, FIGARO, and HEXAGONE balloon experiments show a very
significant annihilation emission component $\approx 5$ $10^\circ$ north of the
galactic plane in the general direction defined by the axis of the galactic
center lobe (Pohl, Reich \& Schlickeiser 1992). Imaging limitations of the
$\gamma$-ray telescopes make it impossible to resolve fine structure, but the
high latitude component appears extended rather than pointlike. If near the
galactic center at distance 8$d_8$ kpc, then e$^+$ sources are producing
$\approx (2-3f/2)^{-1} 4\cdot 10^{42} d_8^2$ $e^+$ s$^{-1}$ which annihilate
$\approx 1$-$2$ kpc above the galactic plane. Here $f$ is the fraction of e$^+$
which annihilate via positronium (Ps) formation. 

Positron sources include radioactive emitters from supernovae and novae, black
hole jets, and low-energy ($\approx 10$-100 MeV nuc$^{-1}$) cosmic rays. Due to
proximity to the galactic center and the feebleness of annihilation radiation 
from star-forming regions, we conclude that the region of enhanced e$^+$-e$^-$
annihilation radiation exists about 1 kpc above the galactic center. A
starburst episode within the inner few hundred pc of our galaxy would drive hot
pair-laden gas into the halo, with the one-sidedness pointing to the site of
initial pressure release at the onset of the starburst activity. Here we
investigate the production, transport, and annihilation of e$^+$ convected to
high galactic latitudes in a wind produced in the galactic center region.

In Section 2, we consider e$^+$ production from supernovae, and summarize 
evidence for starburst activity in the galactic center region.  A model for
e$^+$ transport and annihilation in a fountain model is described in Section
3, and used to calculate a map of the annihilation emissivity from the 
direction of the galactic center.  Predictions and a summary of the results
are given in Section 4.

\section{Positron Production from Supernovae and Galactic Center Activity}

The origin of positrons through the decay of radioactive nuclei is confirmed by
observations of $^{56}$Co and  $^{57}$Co nuclear decay lines from SN 1987A
(Matz et al. 1988; Kurfess et al. 1992), the $^{26}$Al 1.809 MeV line found in
clumped structure along the galactic plane (Diehl et al. 1995), and the
$^{44}$Ti nuclear decay line from Cas A (Iyudin et al. 1994). The mean
lifetimes of the $^{56}$Ni$\rightarrow^{56}$Co and
$^{56}$Co$\rightarrow^{56}$Fe decays in the
$^{56}$Ni$\rightarrow^{56}$Co$\rightarrow^{56}$Fe chain are 8.8 and 111.4 days,
respectively, with a $\beta^+$ emitted 19\% of the time in the latter reaction.
Type Ia supernovae involving white dwarf detonantion or deflagration produce a
time-averaged e$^+$ production rate $\dot N_+^{\rm 56,Ia} \cong 1.4\cdot
10^{43}\eta_{-2} M_{56}\dot N_{{\rm Ia}/{\rm C}}$ e$^+$ s$^{-1}$, where
$M_{56}$ is the average number of Solar masses of synthesized $^{56}$Fe per SN
Ia, $\dot N_{{\rm Ia}/{\rm C}}$ is the number of SN Ia per century throughout
the Milky Way, and $\eta = 10^{-2}\eta_{-2}$ is the escape fraction. Chan \&
Lingenfelter (1993) calculate the escape fraction to be in the range
$0.1\lesssim \eta_{-2} \lesssim 10$, and report (Woosley \& Weaver 1992) that
$0.6< M_{56} < 0.9$.  For core-collapse Type II supernovae, only an average
$0.08 M_\odot$ of $^{56}$Fe is synthesized per SN, and the escape fraction
could reach $0.7$\% for well-mixed ejecta, implying that $\dot N_+^{\rm 56,II}
\cong 10^{42}\eta_{-2} (M_{56}/0.08)\dot N_{II/{\rm C}}$ e$^+$ s$^{-1}$. 

The mean lifetime of the  $^{44}$Ti$\rightarrow^{44}$Sc$\rightarrow^{44}$Ca
chain is 78 yrs, with a positron produced 95\% of the time in the latter decay.
Compared to the mass of $^{56}$Fe, a $^{44}$Ti mass fraction $\zeta =
10^{-4}\zeta_{-4}$ in the range $0.3\lesssim \zeta_{-4}\lesssim 1.4$ is
calculated for SN Ia and $\zeta_{-4}<25$ is found for SN II (Chan \&
Lingenfelter 1993).  This gives $\dot N_+^{44}\cong 8\times
10^{41}\zeta_{-4}M_{56}\dot N_{\rm SN/C}$, noting that most e$^+$ escape from
the SN ejecta and mix with the surrounding medium. $^{26}$Al decays into
$^{26}$Mg with a mean lifetime of $10^6$ yrs, producing a positron 82\% of the
time.  Production of $^{26}$Al is most important in Type II SNe, though only
$0.3\cdot 10^{-5}\lesssim M_{26}\lesssim 20\cdot 10^{-5}$ Solar masses of
$^{26}$Al are produced per SN II, depending on initial stellar mass and the
treatment of semi-convection (Prantzos 1996). This gives $\dot N_+^{26,{\rm
II}}\cong 10^{41} (M_{26}/10^{-5}) \dot N_{\rm II/C}$ e$^+$ s$^{-1}$, a value
probably insufficient to account for the 1.5 M$_\odot$ of $^{26}$Al required to
explain the total observed {\it (11)} 1.809 MeV $^{26}$Al line flux of
$3.1(\pm0.9)\cdot 10^{-4}$ cm$^{-2}$ s$^{-1}$ for which, at least, Wolf-Rayet
stars and novae make important contributions. 

An episode of starburst activity in a region a few hundred pc across enclosing
the galactic center could account for Ginga observations (Koyama et al. 1989;
Yamauchi et al. 1990) of 6.7
keV emission from He-like Fe at temperatures $\gg 10^7$ K.  The hot gas in this
scenario is produced by $10^3$ SNe over the preceeding $10^5$ yrs, which would
be mostly SNe II from massive star evolution unless the activity had persisted
for a period much exceeding $10^5$ yrs.  Other indications (Hartmann 1995)  of
explosive events near the galactic center include large-scale X-ray structures,
radio structures, and indeed the 1.809 MeV line. The radio emission of the
galactic center lobe is explained by the synchrotron emission of nonthermal
electrons convecting and diffusing outward (Pohl et al. 1992). 

Within the limits of uncertainty, $\beta^+$ production from SNe II could
account for the flux of the high-latitude annihilation glow if $\dot N_{\rm
SN/C} \approx 1$ in the central 100-200 pc nuclear region of our galaxy and,
furthermore, if e$^+$ are transported to the annihilation site and efficiently
annihilated.  Pair-laden hot gas from the galactic center starburst would vent
into the galactic halo while expanding, radiating and cooling, and slowing in
transit from the galactic center to the lower-pressure galactic halo. While
convecting outward, e$^+$ would cool and annihilate with the plasma electrons
in the hot wind, and this is our explanation for the observations (Purcell et
al. 1997a, 1997b; Cheng et al. 1997). The venting would be preferentially
one-sided due to the location of the initial starburst activity and the
morphology of the confining gas.\footnote{We note that either Sgr A*, the
putative $10^6$ M$_\odot$ black hole at the Galactic Center, or the less
massive Einstein source (1E 1740.7-2428), weighing in at $\sim 10$-$10^2
M_\odot$, could produce a large enough $e^+$ production rate if annihilation
were efficient, but the origin of the elevated target region requires
additional explanation.} 

\section {Annihilation Fountain Model}

We sketch a model for the fountain of rising annihilating gas energized by a
starburst episode.\footnote{The general picture also applies to plasma ejection
from black hole jets.} Suppose that gas rises with speed $10^7 v_7$ cm s$^{-1}$
from a region of radius $r_b = 100 r_{100}$ pc undergoing a starburst phase,
implying a gas crossing time of order $r_{100}/v_7$ million years.  We
approximate the shape of the volume of the  gas expanding into the galaxy's
halo by an inverted cone with opening angle $\chi$, cut off at both ends, so
that the cross sectional area of the fountain is $\pi r_b^2(1+u)^2$, where
$u=z\tan\chi/r_b$ and $z$ is the height above the galactic plane.  Our major
simplification, which can be relaxed in more general treatments, is that the
gas rises with constant velocity.  Continuity of the mass flux for a
steady-state situation considered here (see Ramaty et al. 1992 for a treatment
of time-dependent injection and annihilation in a uniform medium) implies that
the density of the gas at the base of the fountain is $n_p^0 \cong 0.13 \dot
M_{\odot/{\rm C}}/(v_7r_{100}^2)$ cm$^{-3}$, where $\dot M_{\odot/{\rm C}}$
Solar masses of gas are expelled per century from the starburst region and rise
into the halo. The time- and spatially- averaged density distribution as a
function of height $z$ above the galactic plane is therefore $n_p(z) =
n_p^0/(1+u)^2$. 

The time scale for a positron injected with kinetic energy
m$_e$c$^2$($\gamma-1$) to thermalize with the background hot thermal gas is
controlled primarily by Coulomb losses at mildly relativistic and
nonrelativistic energies.  Coulomb losses operate on a time scale of $\approx
0.5\beta(\gamma - 1)/(n_{-1} \Lambda_{30})$ million years, where $\Lambda_{30}$
is the Coulomb logarithm divided by 30, the density is 0.1$n_{-1}$ protons
cm$^{-3}$, and $\beta c$ is the positron's speed. The kinetic energy
distributions of positrons entering the ISM after SN II explosions are given
for the $^{56}$Co$\rightarrow^{56}$Fe and the $^{44}$Sc$\rightarrow^{44}$Ca
decays by functions peaking near 0.6 MeV with FWHM widths of $\approx 0.9$ MeV
and high-energy tails reaching to $\approx 1.45$ MeV (Chan \& Lingenfelter
1993).  The fraction of e$^+$ which annihilate in flight  prior to thermalizing
usually amounts to $< 10$\% and these positrons do not contribute to the 0.511
line emission (Murphy, Dermer \& Ramaty 1987).  After entering the thermal
pool, the positrons annihilate in a fully ionized thermal plasma through direct
annihilation and radiative combination on a thermal annihilation time scale of
$30/(\lambda_{-14}n_{-1})$ million years, where the annihilation rate
coefficient $\lambda = 10^{-14}\lambda_{-14}$ cm$^3$ s$^{-1}$ (Bussard, Ramaty
\& Drachman 1979). When a significant fraction of neutral atoms or partially
ionized ions are present, the annihilation rate can dramatically increase
because $\lambda_{-14}\rightarrow 10^6 $ near temperatures of $\simeq 10^5$ K
due to the onset of charge exchange processes which have atomic-sized cross
sections.  A pair of 0.511 MeV line $\gamma$ rays comes from direct
annihilation in the thermal gas and, one-quarter of the time, from annihilation
via Ps formation. 

We calculate the annihilation flux at height $z$ after tracking the energy
evolution and spatial propagation of the e$^+$ following injection.  The e$^+$
height-dependent injection function is approximated by a Gaussian function
which is allowed to be, in general, offset from the galactic plane.  The FWHM
width of the injection function is taken to be 180 pc, comparable to twice the
scale height of massive stars. Though diffusion must be important for very
relativistic e$^+$ (Lerche \& Schlickeiser 1980), we assume that the nonthermal
and thermal e$^+$ are entrained in the hot gas and convect away from the
galactic midplane with constant speed $v_0$.  The general e$^+$ equation of
motion involves Coulomb, bremsstrahlung, adiabatic expansion, synchrotron and
Compton energy losses, but for $\beta^+$ injection, only Coulomb and adiabatic
expansion losses are important.  The e$^+$ energy-loss rate from adiabatic
expansion is given by $-\dot\gamma_{\rm adia} = (4\gamma-3-\gamma^{-1})\dot
V/(3V)$, which bridges the nonrelativistic and relativistic regimes and applies
to high-beta plasmas, which is suitable for magnetic fields weaker than $\sim
10^{-3}$ Gauss. The volume expansion rate for the fountain geometry is $\dot V/
V = 2 v_0\tan\chi/[r_b(1+u)]$, from which we find that Coulomb losses dominate
adiabatic expansion losses when $\beta(4\gamma-3-\gamma^{-1}) \lesssim K =
36\Lambda_{30}\dot M_{\odot/{\rm C}}/[v_7^2 r_{100}(\tan\chi/0.1)]$. 

After $\beta^+$ emission, positrons convect away from the galactic plane and,
if injected with sufficiently low energies, thermalize with the background gas
through Coulomb processes (excepting those few which annihilate in flight).  We
call the distance between injection and thermalization the Maxwell-Boltzmann
length (MBL), which is a nonlinear function of the injection height $z_i$ and
the e$^+$ injection kinetic energy m$_e$c$^2(\gamma - 1)$.  When Coulomb
processes dominate, the MBL is given by $u_{\rm MB} = [(1+u_i)^{-1} -
K_c^{-1}(\beta_i\gamma_i - \arccos \gamma_i^{-1})]^{-1} - 1$, where $u_{\rm
MB(i)} = z_{\rm MB(i)}\tan\chi/r_b$.  The constant $K_c = 2.4 \dot
M_{\odot/{\rm C}}\Lambda_{30}/(v_7^2 r_{100}\tan\chi)$. 

When $K\gg 1$, as with standard parameter values, positrons injected through
$\beta^+$ production thermalize close to their injection site. Thus the spatial
dependence of positrons thermalizing with hot gas is, in this regime, equal to
the energy-integrated $\beta^+$-injection function.  Following thermalization,
a positron continues to convect upward into the galactic halo until it either
annihilates or merges with the dilute halo gas.  The decay law through
annihilation for the rising positrons is governed by the value of the
temperature-dependent reaction rate coefficient $\lambda (T)$, the ionization
state, composition, and density of the medium. For hot gases with $T \gtrsim
10^6$ K, the 2-photon direct annihilation channel is most important with
$\lambda_{-14} \cong 1$.  The decay law for thermal positrons annihilating in a
thermal gas is $-\dot N_+(t) \cong N_+(t) \lambda n_p(t)$. From this equation,
we derive the height-dependent differential production rate of 0.511 MeV
annihilation line photons, given by 
\begin{equation}
\dot N_{\rm 0.511~MeV}(z) = {2n_-^0 \lambda(T)\over (1+u)^2 v_0} \int_0^z
dz^\prime\;\dot N^{\rm th}_+(z^\prime)\;\exp [-K_a({1\over 1+u^\prime} -
{1\over 1+u})]\;. 
\end{equation}
Here $\dot N_+^{\rm th}(z)$ is the MB injection function differential in height
$z$, which is  obtained by convolving the $\beta^+$ injection function with its
MBL and integrating over the initial energies of the $\beta^+$ positrons. The
constant $K_a =0.039 \lambda_{-14}\dot M_{\odot/{\rm C}}/(v_7^2
r_{100}\tan\chi)$, $n_-^0$ is the electron density at the base of the fountain,
and $u^{(\prime)} = z^{(\prime)}\tan\chi/r_b$. 

In the approximation where e$^+$ thermalize close to their injection site,
equation (1) is easily solved to give the results shown in Fig. 1. The total
e$^+$ injection rate is $10^{42}\dot N_{+42}$ e$^+$ s$^{-1}$ with $\dot
N_{+42}= 1$. The solid curves give the integral 0.511 MeV photon production
rate between the galactic midplane and height $z$, and the dashed curves give
the differential 0.511 MeV production rate in units of 0.511 MeV ph s$^{-1}$
z(pc)$^{-1}$.  The dotted curves represent the spatial e$^+$ distribution
function, which are centered at and 50 pc above the galactic midplane in Figs.
1a and 1b, respectively. 

Before interpreting Fig. 1, note that $\dot N_{\rm
0.511~MeV}(<\infty)\rightarrow 2\cdot 10^{42}$ 0.511 MeV ph s$^{-1}$ if all
$e^+$ annihilate.  This limiting value is reached only if most $\beta^+$
injection occurs high above the galactic plane so that few e$^+$ are convected
to negative values of $z$. Standard parameter assignment with no source offset
and $\tan\chi = 0.1$ (Fig. 1a), corresponding to a 6$^\circ$ fountain opening
angle, yields an integral annihilation flux of $\approx 3\cdot 10^{41}$ 0.511
MeV ph s$^{-1}$, implying a one-sided annihilation efficiency of $\sim 15$\%.
In contrast, when the fountain opens to 45$^\circ$ ($\tan\chi = 1$), the
annihilation efficiency plummets because of the severe thinning of gas density
with height. In Fig. 1b, the wind speed parameter for the hot gas is increased
by an order of magnitude from the standard value and the injection Gaussian is
offset by  50 pc. This also reduces the annihilation efficiency because flux
continuity demands a reduced density in a steady-state approximation. 

The overall shape of the differential 0.511 MeV production function is unusual
in that it exhibits a broad plateau between $\approx 50$ pc and 1-2 kpc. Rather
than decreasing monotonically with height above the galactic midplane, the
maximum of the differential 0.511 MeV annihilation flux peaks downstream from
the maximum of the Gaussian injection function.  This can be understood by
superposing contributions from discrete sources. The annihilation flux
decreases downstream from a steadily emitting $e^+$ source as the radioactive
debris convect outward with constant velocity. But the divergence of the wind
vector at the galactic midplane means that the e$^+$ density at a height $z_0$ 
is proportional to the number of sources between the galactic midplane and
$z_0$. This quantity increases with height above the galactic midplane for
reasonable injection distributions until the source injection function tails
off.  The underlying assumption of time- and spatial-averaging appears to be
satisfied for our system, because $\sim 10^4$ SNe occur during the gas-crossing
time scale of the starburst region. More detailed calculations must consider
the spatial distribution and temporal evolution of the discrete SN events
exploding throughout the region. 

The rate at which 0.511 MeV annihilation photons are emitted between $\approx
0.1$ and 2 kpc is $\cong 2\cdot 10^{41}\dot N_{+42}$ and $\cong 3\cdot
10^{41}\dot N_{+42}$ 0.511 MeV ph s$^{-1}$ for the top curves in Figs. 1a and
1b, respectively.  To match the OSSE/{\it SMM}/TGRS/balloon data analyses 
(Purcell et al. 1997a, 1997b; Cheng et al. 1997)
implies that annihilation occurs at the rate of $\approx 4\cdot
10^{42} d_8^2$ 0.511 MeV ph s$^{-1}$.  Thus injection rates of $\dot N_{+42}
\simeq 20$ and$\dot N_{+42} \simeq 13$ are required to match the observations
for these two cases.  Such rates could be supplied by $\beta^+$ production from
freshly synthesized $^{44}$Ti in SN II given the uncertainty of the $^{44}$Ti
mass fraction, the SNe rate, and the parameters of the outflowing wind. This
rate could perhaps also derive from $\beta^+$ production in the
$^{56}$Ni$\rightarrow^{56}$Co$\rightarrow^{56}$Fe chain if either the
calculated  escape fraction or amount of synthesized $^{56}$Ni per SN II were
underestimated.  Black hole e$^+$ production could provide an additional source
of the positrons. 

Fig. 2 shows a contour map of the annihilation flux produced by the galactic
fountain, added to a disk and galactic bulge distributions fitted to the
measured (Purcell et al. 1997a) annihilation emissivity. The axis of the
fountain is directed away from the center of our galaxy along the direction of
the galactic center lobe, which is inclined in projection by $\sim 20^\circ$
from the axis of the North Galactic Pole. (Note that the location of the
centroid of the spheroidal bulge flux is offset from the location of the
galactic center.) The agreement of this idealized model with the data is
satisfactory, and predicts that peak enhancement of the fountain's annihilation
flux occurs $\gtrsim 100$ pc above the galactic plane. 

\section{Predictions and Summary}

Due in large part to gamma-ray observations with the OSSE instrument on {\it
Compton}, a new component of the ISM has been discovered: hot plasma pushing
into the galactic halo from  a region of starburst activity surrounding the
galactic center a few hundred pc in extent. The existence of the outflowing
wind is suggested by radio and X-ray observations (Pohl et al. 1992; Koyama et
al. 1989; Yamauchi et al. 1990; Morris \& Serabyn 1996) and represents a
low-power analog of nuclear starburst outflows detected in M82 and NGC 253
(Shopbell \& Bland-Hawthorn 1997). Hot ionized plasma can be detected through
pulsar dispersion measurements, and we predict a dispersion measure jump of
$\sim 50$ when scanning across the fountain several degrees above the galactic
plane. The difficulty is to have independent distance measurements of pulsars
located $>8$ kpc away. Continuum free-free and recombination Ly$\alpha$ lines
of H, He, and Ps are predicted from the fountain, but obscuring foreground gas
makes it difficult to view the galactic center region at optical and UV
wavelengths.  As $^{26}$Al is convected upward with the flow, a characteristic
decay law for the diffuse 1.809 MeV emission (Chen, Gehrels \& Diehl 1995)
should be observed, though at a flux level requiring at least {\it INTEGRAL}
telescope sensitivities. Turbulent hot gas in the central region might also
explain the broadened 1.809 MeV lines observed (Naya et al. 1996) with the GRIS
instrument. The width of the 2$\gamma$ 0.511 MeV line from annihilation in hot
gas will be broader than the galactic disk 0.511 MeV line emission because of
the larger temperature of the plasma, depending in detail on the dust content
of the fountain, and the 3$\gamma$ Ps continuum fraction $f$ will be spatially
varying (Ramaty, private communication). 

If the annihilation fountain is our first clear view of a conduit of hot gas
venting the starburst activity near the galactic center into the galactic halo,
then it is likely that cosmic rays are efficiently transported into and diffuse
throughout the halo along this route.  The $>100$ MeV EGRET (Hunter et al.
1997) and 408 MHz Effelsberg (Haslam et al. 1982) all-sky maps show enhanced
emission north of the galactic plane which could be ascribed to this effect.
Positrons which fail to annihilate in the fountain will diffuse throughout the
halo and annihilate if they encounter the gaseous disk. This could explain the
large scale height of the disk component of the diffuse annihilation radiation
$\gtrsim 10$-20$^\circ$ away from the galactic center (Ramaty et al. 1994).
Long-lived radioactivity produced by SNe would also trace this channel.
Small-scale variations of the 0.511 MeV map would indicate localized OB
associations, changes in the ionization state of the gas, and sites of black
hole e$^+$ injection.  More observations and analyses are required to
understand better the activity occurring near the center of the Milky Way. 

\acknowledgments

We thank members of the OSSE team, especially W. Purcell and J. Kurfess, for
conversations about the annihilation radiation observations.  We also
acknowledge useful discussions with Drs. R. Ramaty, R. Schlickeiser, J. Lazio,
and A. Harding. In addition, we acknowlege support from the Office of 
Naval Research.

\clearpage
\begin{figure*}
\vskip -3truecm
\centerline{\psfig{figure=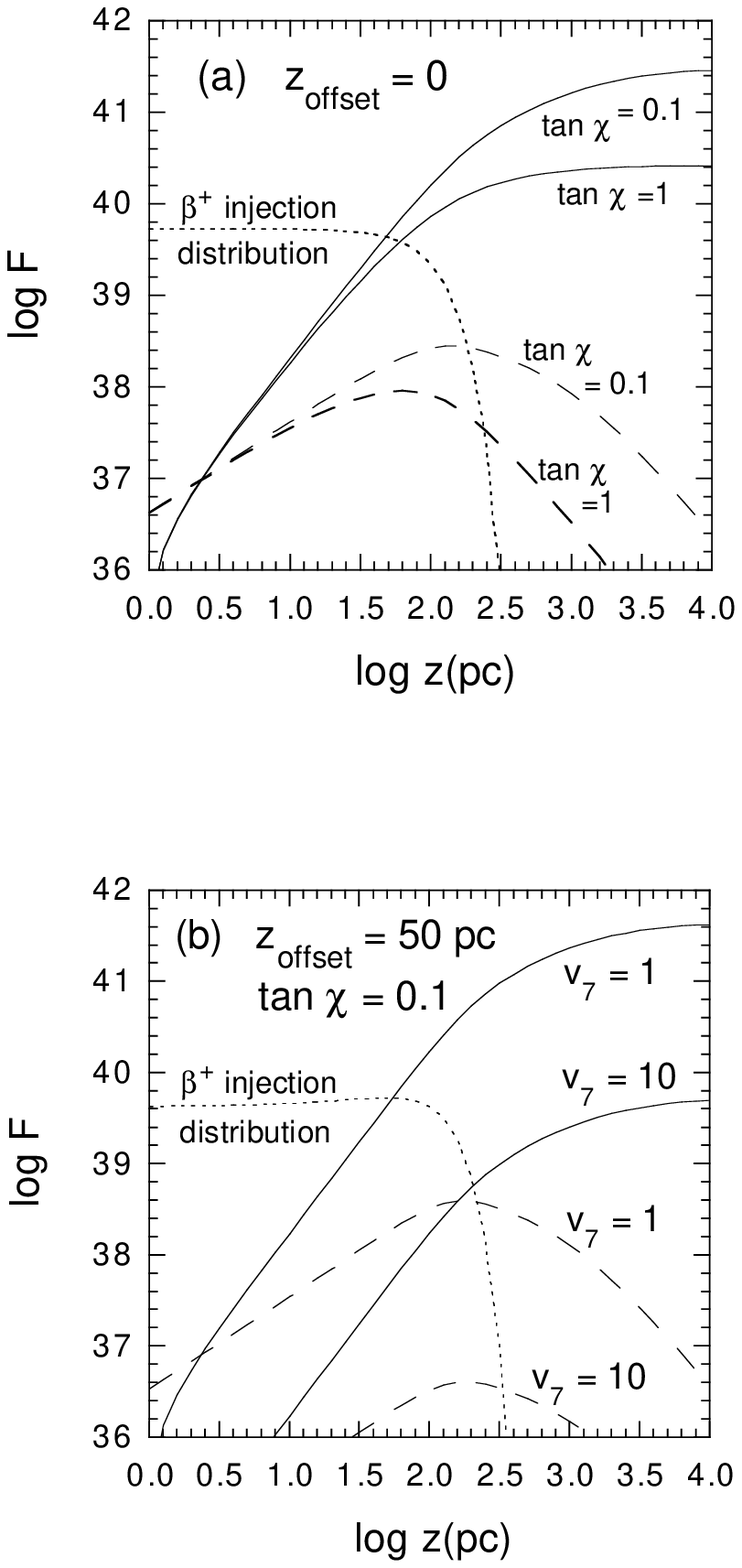,height=24 cm,width=20 cm}}
\vskip -4truecm
\caption{ \em
Height distribution of the differential (dashed
curves) and integral (solid curves) 0.511 MeV annihilation emissivities
produced in a fountain of hot gas rising upward with constant velocity $v_0$. 
Except where noted, the opening angle of the fountain is $\chi = \arctan (0.1)
= 5.^\circ 7$, $v_0 = 10^7 v_7$ cm s$^{-1}$ with $v_7 = 1$, and the radius of
the base of the fountain is 100 pc.  We assume that 1 Solar mass of hot gas is
ejected into the fountain per century.  The dotted curves show the $\beta^+$
spatial injection function normalized to a total injection rate of
$10^{42}$ e$^+$ s$^{-1}$. Fig. 1(a) compares the effects of different opening
angles $\chi$ for an injection function symmetric about the galactic midplane. 
Fig. 1(b) compares the effect of varying gas speeds for an injection function
offset from the galactic midplane by 50 pc. }
\end{figure*}

\begin{figure*}
\centerline{\psfig{figure=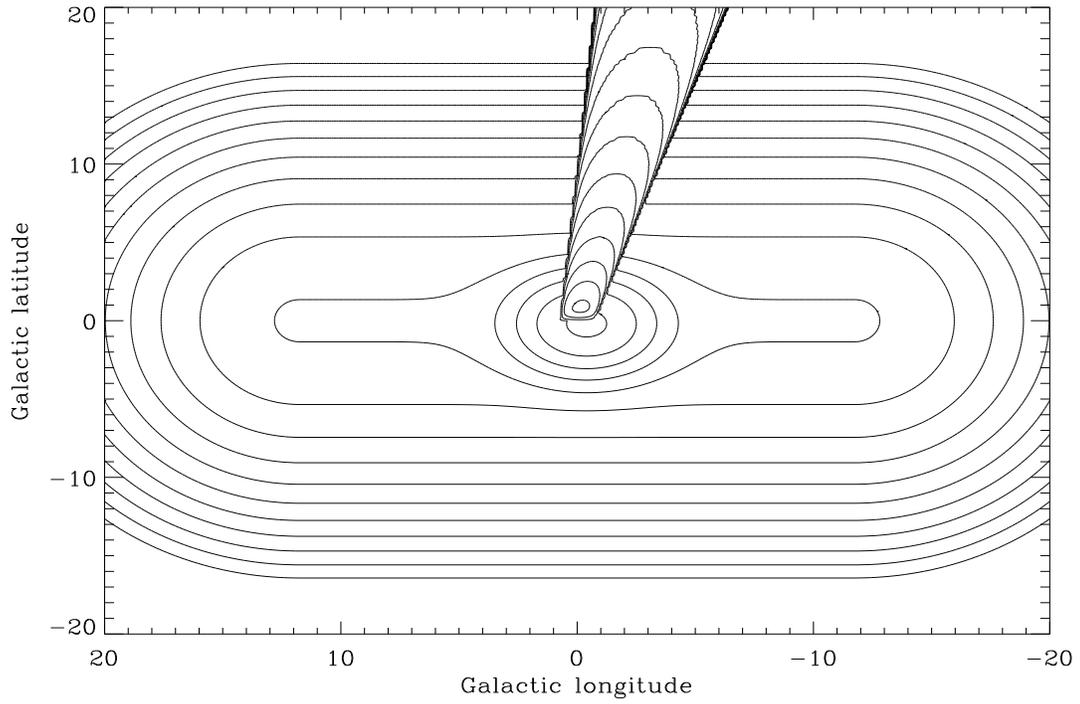,height=10 cm,width=15 cm}}
\vskip 0.8truecm
\caption{ \em
Contour plot of the model annihilation emissivity of
the Milky Way including the disk, galactic bulge, and fountain component.  The
contours are in units of $10^{-(2+n)/5}$ 0.511 MeV ph cm$^{-1}$ s$^{-1}$
sr$^{-1}$, with the central contour corresponding to $n = 1$. }
\end{figure*}

\end{document}